# A COMPOSITE DESIGN PATTERN FOR SERVICE INJECTION AND COMPOSITION OF WEB SERVICES FOR PEER-TO-PEER COMPUTING WITH SERVICE-ORIENTED ARCHITECTURE

Vishnuvardhan Mannava[1] and T. Ramesh[2]

[1]Department of Computer Science and Engineering, K L University,
Vaddeswaram, 522502, A.P., India
`vishnu@kluniversity.in`
[2]Department of Computer Science and Engineering, National Institute of Technology,
Warangal, 506004, A.P., India
`rmesht@nitw.ac.in`

*ABSTRACT*

*In this paper we present a Service Injection and composition Design Pattern for Unstructured Peer-to-Peer networks, which is designed with Aspect-oriented design patterns, and amalgamation of the Strategy, Worker Object, and Check-List Design Patterns used to design the Self-Adaptive Systems. It will apply self-reconfiguration planes dynamically without the interruption or intervention of the administrator for handling service failures at the servers. When a client requests for a complex service, Service Composition should be done to fulfil the request. If a service is not available in the memory, it will be injected as Aspectual Feature Module code. We used Service Oriented Architecture (SOA) with Web Services in Java to Implement the composite Design Pattern. As far as we know, there are no studies on composition of design patterns for Peer-to-peer computing domain. The pattern is described using a java-like notation for the classes and interfaces. A simple UML class and Sequence diagrams are depicted.*

## 1. INTRODUCTION

The most widely focused elements of the autonomic computing systems are self-* properties. So for a system to be self-manageable they should be self-configuring, self-healing, self-optimizing, self-protecting and they have to exhibit self-awareness, self-situation and self-monitoring properties [2]. As the web continues to grow in terms of content and the number of connected devices, peer-to-peer computing is becoming increasingly prevalent. Some of the popular examples are file sharing, distributed computing, and instant messenger services. Each one of them provides different services, but shares the same mechanism like Discovery of peers, searching, file and data transfer. Currently developed peer-to-peer applications are inefficient with the developers solving the same problems and duplicating the similar infrastructure implementations [15]. Most of the applications are specific to a single platform and can't communicate and share data with different applications.

DOI : 10.5121/ijwsc.2012.3305 49

To overcome the current existing problems Sun Microsystems have introduced JXTA. JXTA is an open set, generalized peer-to-peer (P2P) protocols that allows any networked device sensors, cell phones, PDA's, laptops, workstations, servers and supercomputers- to communicate and collaborate mutually as peers. The advantage of using the JXTA peer-to-peer programming is that it provides protocols that are programming language independent, multiple implementations, know as bindings, for different environments. The JXTA protocols are all fully interoperable. So with help of JXTA programming technology, we can write and deploy the peer-to-peer services and applications. JXTA protocols standardize the manner in which peers will discover each other, self-organize into peer groups, Advertise and discover network resources, communicate with each other, monitor other.

JXTA overcomes the many of the problems in current existing peer-to-peer systems, some of them are: 1) Interoperability enables the peers provisioning P2P services to locate and communicate with one another independent of network addressing and physical protocols. 2) Platform Independent-JXTA provides the developing code with independent form programming languages, network transport protocols, and deployment platforms. 3) Ubiquity JXTA is designed to be accessed by any device not just the PC or a specific deployment platform. In this paper we propose a design pattern for providing the services to peer-clients in unstructured peer-to-peer network.

Design patterns are most often used in developing the software system to implement variable and reusable software with object oriented programming (OOP) [6]. Pattern composition has been shown as a challenge to applying design patterns in real software systems. Composite patterns represent micro architectures that when glued together could create an entire software architecture. Thus pattern composition can lead to ready-made architectures from which only instantiation would be required to build robust implementations. A composite design patterns shows a synergy that makes the composition more than just the sum of its parts. As far as we know, there are no studies on composition of design patterns for Peer-to-peer computing domain. Most of the design patterns in [6] have been successfully applied in OOPs, but at the same time developers have faced some problems like as said in [10] they observed the lack of modularity, composability and reusability in respective object oriented designs [7]. They traced this lack due to the presence of crosscutting concerns. Crosscutting concerns are the design and implementation problems that result in code tangling, scattering, and replication of code when software is decomposed along one dimension [16] e.g., the decomposition into classes and objects in OOP. To overcome this problem some advanced modularization techniques are introduced such as Aspect-oriented programming (AOP) [15] and Feature-oriented programming (FOP). In AOP the crosscutting concerns are handled in separate modules known as aspects, and FOP is used to provide the modularization in terms of feature refinements.

In our proposal of a design pattern for a peer-to-peer system, we use the Aspect-oriented design pattern called Worker Object pattern [14] and Checklisting Design Pattern [5]. When comes to the worker object pattern it is an instance of a class that encapsulates a method called a worker method. It will create and handle each client service request in separate thread by making the job of server easy from looking after each and every client until it completes serving its request .So the server can listen for new client requests if any to handle. The Checklisting Design Pattern is used to provide means for selecting a service plan that best suits the clients request and also that matches the WSDL information of a Service providing peer server. An application has to interact with considering some constraints. We come across this situation when a task needs to be accomplished by the collaboration of multiple peers. The strategy Design Pattern helps in making decision regarding which Checklist of items (items here are the services) have to be selected to solve the complex task requested by user with respect to the WSDL information available from each and every service providing peer servers [9].

## 2. RELATED WORK

In this section we present some works that deal with unstructured peer-to-peer systems design. There are number of publications representing the design pattern oriented design of the peer-to-peer computing systems. The JXTA protocols standardization provides one of the autonomic computing system properties known as self-organization into peer groups. The self-organization is property that provides the autonomic capability in the peer-to-peer design of networks.
In V.S.Prasad Vasireddy, Vishnuvardhan Mannava, and T. Ramesh paper [12] discuss applying an Autonomic Design Pattern which is an amalgamation of chain of responsibility and visitor patterns that can be used to analyze or design self-adaptive systems. They harvested this pattern and applied it on unstructured peer to peer networks and Web services environments.

In Sven Apel, Thomas Leich, and Gunter Saake [1] they proposed the symbiosis of FOP and AOP and aspectual feature modules (AFMs), a programming technique that integrates feature modules and aspects. They provide a set of tools that support implementing AFMs on top of Java and C++.
In Alois Ferscha, Manfred Hechinger, Rene Mayrhofer, Ekaterina Chtcherbina, Marquart Franz, Marcos dos Santos Rocha, Andreas Zeidler [5] they proposed that The design principles of pervasive computing software architectures are widely driven by the need for opportunistic interaction among distributed, mobile and heterogeneous entities in the absence of global knowledge and naming conventions. Peer-to-Peer (P2P) frameworks have evolved, abstracting the access to shared, while distributed information. To bridge the architectural gap between P2P applications and P2P frameworks we propose patterns as an organizational schema for P2P based software systems. Our Peer-it hardware platform is used to demonstrate an application in the domain of flexible manufacturing systems.

In Vishnuvardhan Mannava, and T. Ramesh paper [17] they have proposed a design pattern for Autonomic Computing System which is designed with Aspect-oriented design patterns and they have also focused on the amalgamation of the Feature-oriented and Aspect-oriented software development methodology and its usage in developing a self-reconfigurable adaptive system.
In Vishnuvardhan Mannava, and T. Ramesh paper [18] they have proposed a system for dynamically configuring communication services. Server will invoke and manage services based on time stamp of service. The system will reduce work load of sever all services in executed by different threads based on time services are executed, suspended and resumed.

In Vishnuvardhan Mannava, and T. Ramesh paper [19] they have proposed an adaptive reconfiguration compliance pattern for autonomic computing systems that can propose the reconfiguration rules and can learn new rules at runtime.

Because of the previous proposed works as described above, we got the inspiration to apply the aspect oriented design patterns along with inclusion of the feature-oriented software development capability to autonomic systems.

## 3. PROPOSED AUTONOMIC DESIGN PATTERN

One of the objects of this paper is to apply the Aspect-oriented design patterns to the object-oriented code in the current existing application. So that a more efficient approach to maintain the system and providing reliable services to the client requests can be achieved.

The very important capability that our proposed design pattern provides in the peer-to-peer computing systems that, when a new service is to be added to the peer system in the network without disturbing the running server we can do this with the help of Aspectual Feature Module [1] oriented insertion of the new service into the peer-server code by using the feature refinements property of Feature-Oriented Programming (FOP). In our proposed Design Pattern for an Autonomic Computing System, initially all the peers in the network group will broadcast the advertisements that represents the respective services that are provided by them as the WSDL messages. The WSDL messages are the XML files that are used in web services. It defines the input and output parameters of a web service in terms of XML schema. With these sent advertisements all the peers will know the information about the services that are provided by different peer servers and whether they are Active or Deactivated at present. Here in each of the Web Service Description Language (WSDL) file it will include the status of the service whether it is active or deactivated currently.

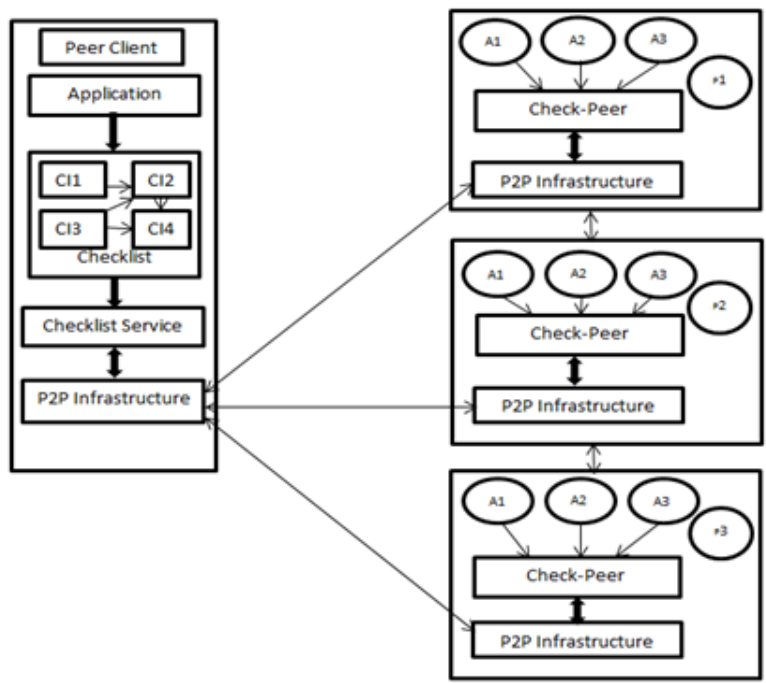

Figure 1: The Proposed Design Pattern for Service Injection and Composition of Web Services for Unstructured Peer-to-Peer networks with SOA.

The Client who requests for a Complex service that cannot be fulfilled with single service but has to take help of two or more services that are provided by different peers in order to satisfy the client's request. So with the help of the Checklist Design Pattern in [5] will help us in selecting the perfect combination of services that best suits the clients service request with respect to the WSDL messages send by the peer servers. Here the Strategy Design Pattern will select a CheckList Plan that better suits the Service Request to be fulfilled and also it will select the perfect WSDL of a peer server that can perfectly process the service request of a Client. For a clear idea of the pattern structure proposed see Figure 1 referred from [5]. Action means the Service definition, Check peer will check whether it can apply the input data retrieved to this service.P2P Infrastructure means any middleware technology like JXTA for establishment of Connections between clients and peers. Once the information of all the peers that are providing

52

the web services is gathered as WSDLs from respective peer servers then it (Client) will invoke the respective peer's service that is responsible for the performing the task.

Here the Peer Web Service provider can take help of the services that are provided by the other peers in the network to fulfil the complex service request of a client. So this kind of service composition can be achieved at the server side with Service Oriented Architecture (SOA) implementation with Web Services. With this invocation call the respective peers will check whether the requested service is currently Active and running. If this is not the case then it will loads the requested service into the memory as a new Feature with the help of the Aspectual Feature Module [1] code from the Service Repository without disturbing the already running services in that peer. Now once the service is loaded into the memory then the peer will invoke a call that will update the State of the service to Activated.

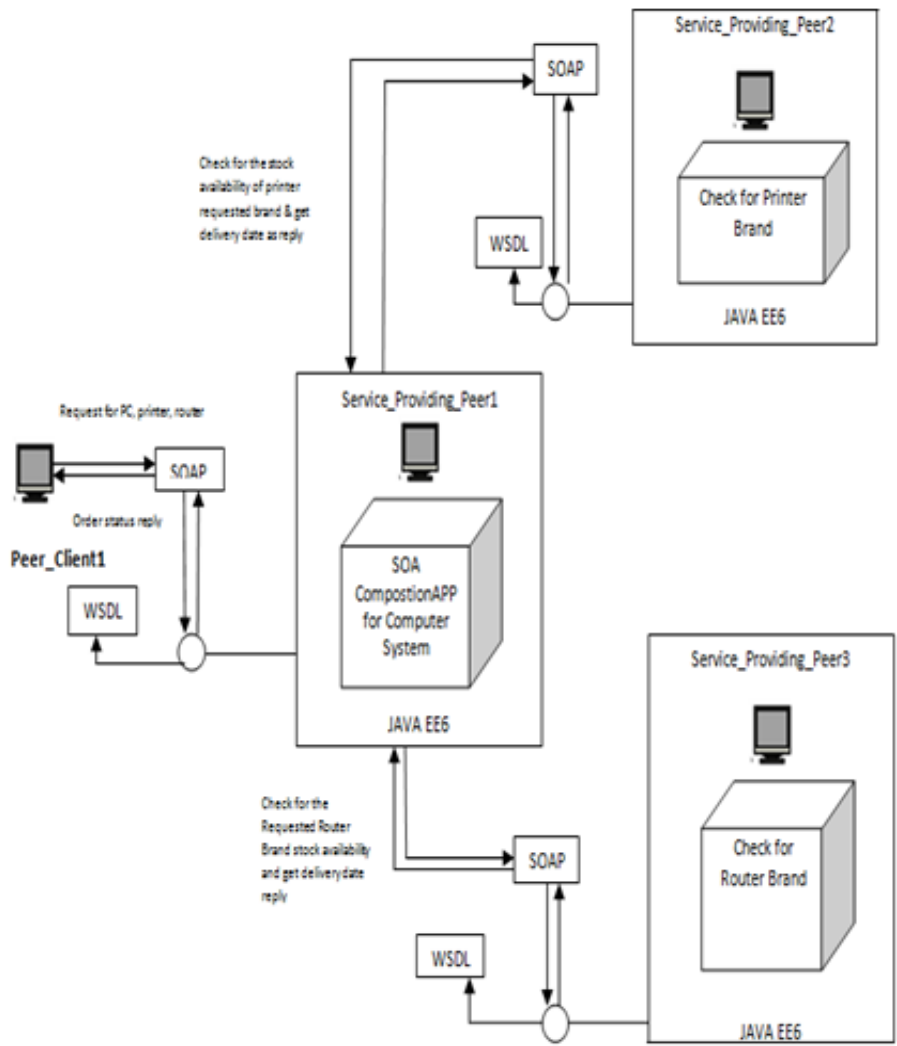

Figure 2: Composition of the Services with Service Oriented Architecture using web services.

The Worker Object Pattern [14] is responsible for handling the different client requests in terms of separate Thread based requests. Means when more than one client request for same service or different one they are handled in different Threads.

So in our design pattern we have provided the capabilities of Web Services Composition[4], Service Invocation, and Inclusion of the new Service Operation as a Feature Module, all the three are provided in each and every server in a Distributed Application.

## 4. DESIGN PATTERN TEMPLATE:

To facilitate the organization, understanding, and application of the proposed design patterns, this paper uses a template similar in style to that used in [13]. Likewise, the Implementation and Sample Code fields are too application-specific for the Design Patterns presented in this paper.

### 4.1. Pattern Name:

Peer-to-Peer service injection design pattern along with composition of services with SOA.

### 4.2. Classification:

Structural –Decision Making

### 4.3. Intent:

Systematically applies the Aspect-Oriented Design Patterns to an unstructured peer-to-peer Computing System and service injection with a Refinement class for providing new service in the peer in terms of a Feature Module.

### 4.4. Context:

Our design pattern may be used when:

- The service requested is a complex service and need to be executed with the help of collaboration of peers in the network.
- To include the new service operations into peers as Aspectual Feature Modules [1].
- For providing the Distributed service request processing environment with the help of JXTA peer-to-peer Programming and Service Oriented Architecture (SOA) with Web Services.

### 4.5. Proposed Design Pattern Structure:

A UML class diagram for the proposed design Pattern can be found in Figure 3.

### 4.6. Participants:

   a) **Client:** This application creates JxtaSocket and attempts to connect to JxtaServerSocket. Peer Group will create a default net peer group and a socket is used to connect to JxtaServerSocket. After these steps the client will call the run method to establish a connection to receive and send data. The startJxta method is called in the client to create a configuration from a default configuration and then instantiates the JXTA platform and

54

creates the default net peer group. Once the net peer group is created the client will send a Pipe Advertisement for requesting a service in the peer-to-peer network.

b) **Checklist Pattern [5]:** In this class it will consists of Checklists that a set of Check items. Here it will provide the service execution plans for the client's requested service.}

c) **Checkpoint Selection:** Here the strategy pattern will help to select the perfect Checklist that matches for the execution of the requested complex service. The selection is based on some rule based selection statements and also that matches the input and output parameter details in WSDL of a particular Peer Server.

d) **Service Providing PeerN (N=1 or 2 or 3):** First the default net peer group is created with Peer Group. Creates a JxtaSeverSocket to accept connections, and then executes the run method to accept connections from clients and send receive data. The peer will start listening for the service requests. If a service that is requested by a client is not running in the memory then the peer will load the new service from the service repository into the memory space of the currently running services and at the same time it will invoke a functional call that will change the state of that service to Activated from deactivated state.

e) **Connection Handler:** This will take care of the connections with multiple clients and sending and receiving the data between the clients and service providing peer.

f) **Worker Object Aspect:** Once the connection is established between the peer and client, the worker object will create a separate thread for the connected client to run the requested service in a new thread for that it will call the Reactor Pattern method to handle the requested service execution.}

g) **Service Repository:** It will contain the services that are currently deactivated and that are not loaded into the memory for execution. The services not being requested by any of the clients are kept here.}

h) **Refines Class Service:** This will add a new service from the service repository that is requested by a client, in such a way that the insertion will be done as a new feature with the help of FOP [3].

The view of our proposed design pattern for the Distributed Computing System can be seen in the form of a class diagram see Figure 1.

The flow of control in the Distributed Computing System can be shown with a sequence diagram in Figure 2.

### 4.7. Consequences:

a) With the use of this pattern we can get the benefits of worker object pattern, The application will handle each of the client requests in a separate thread by reducing the overhead on the main peer thread that is providing the service to get blocked until the first client request is served and making other client requests to get blocked.

b) We use the Feature-Oriented Programming [3] to insert the new service into the memory from the Service Repository as a feature into the current executing services code.

c) By the use of dynamic crosscutting concerns of the Aspect-Oriented Programming [14] the system will me executing fast as the decisions are made at run-time. With the help of the Web Service Description Language (WSDL) all the clients can get information about the services that are active and currently provided by the peer servers

### 4.8. Related Design Patterns:

a) **Case-Based Reasoning Design Pattern [13]:** The Case-Based Reasoning Design Pattern will apply the rule based decision making mechanism to determine a correct reconfiguration plan. This design pattern will separate the decision-making logic from the functional logic of the Application. In our proposed pattern we will use this pattern to provide the perfect suitable plan to implement the customer requested service.

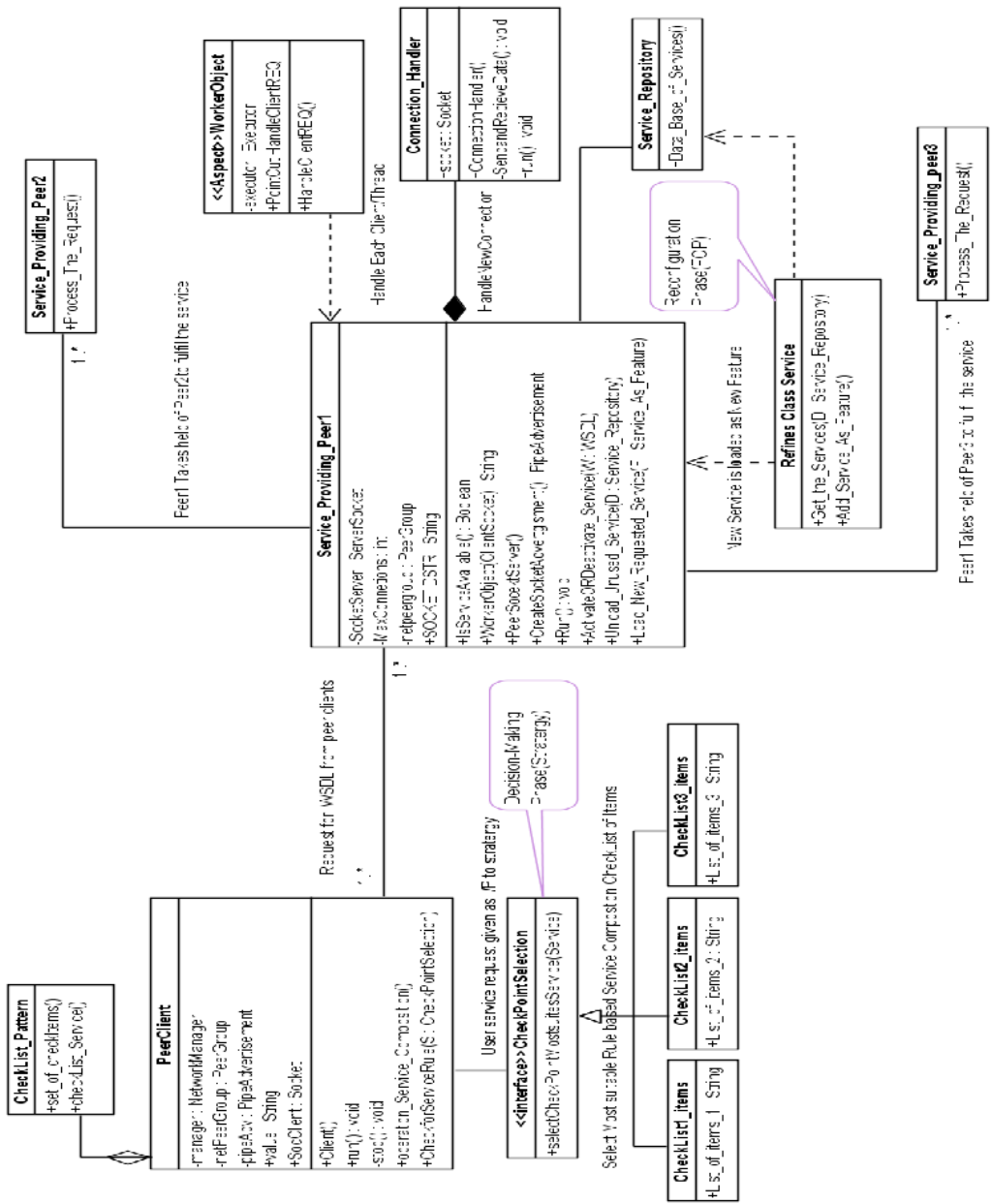

Figure 3: Applying Design Pattern for the peer-to-peer computing System.

### 4.9. Role of our design pattern in Autonomic computing system

**Worker Object Pattern [14]:** The worker object pattern is an instance of a class that encapsulates a worker method. A worker object can be passed around, stored, and invoked. The worker object pattern offers a new opportunity to deal with otherwise complex problems. It will provide the server with the facility to handle the service request form different clients in a separate per client connection. We may use this pattern in different situations like, implementing

57

thread safety in swing applications and improving the responsiveness of the UI applications to performing authorization and transaction management.

**Strategy Design Pattern [6] [8] [11]:** This pattern can be used to define a family of algorithms, encapsulate each one, and make them interchangeable. Strategy lets the algorithm vary independently from the clients that use it.}

**Checklisting Design Pattern [5]:** The Checklisting pattern provides means for maintaining a (usually ordered) list of services (usually running on different peers) an application has to interact with, considering arbitrary constraints. This problem typically arises in mobile scenarios where a task is accomplished by the collaboration of multiple peers.

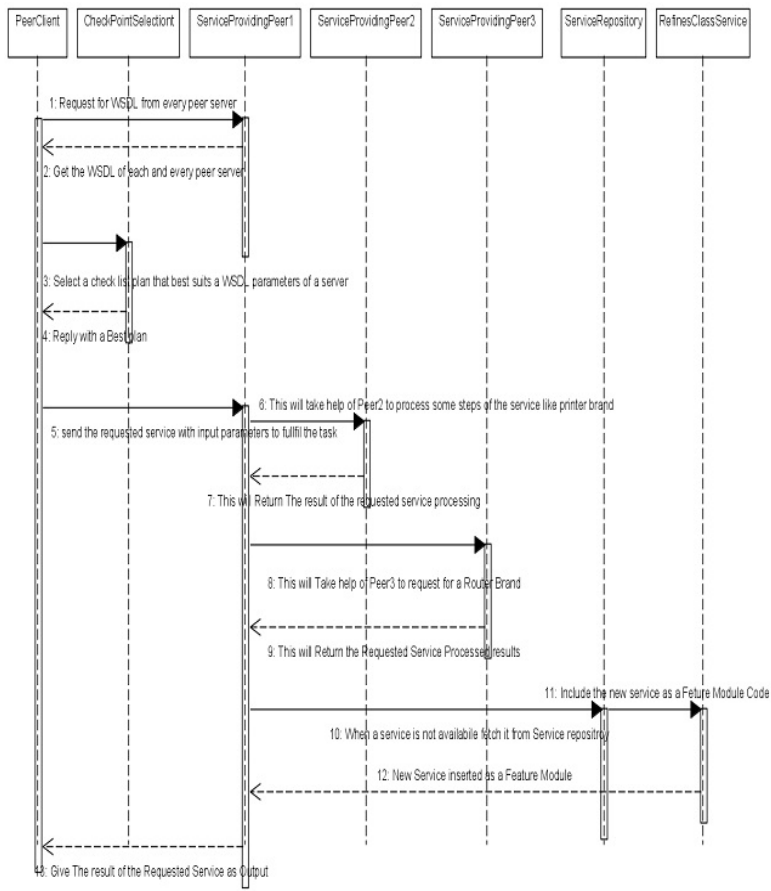

Figure 4: Sequence Diagram for the proposed Design Pattern.

## 5. COMPOSITION OF WEB SERVICES WITH SERVICE ORIENTED ARCHITECTURE (SOA)

Composition of web services is the most important technique that the developers are now a day's most interested in providing the web services with. So in order to achieve the composition in our paper we are adopting to the Service Oriented Architecture (SOA). SOA provides composition of the services provided by the different servers to fulfill the complex service requests of the clients.

We can use the either Java or .Net platform to develop the SOA with web services. Even if you use any one of them it's the web services that make the SOA compatible with the composition of the services in any platform. When using the web services, interface definitions are provided using the WSDL. Each service at the server is associated with a WSDL document. WSDL is expressed using XML. It defines the input and output parameters of a Web service in terms of XML Schema. So with the help of the Simple Object Access Protocol (SOAP) is used as XML-based protocol for exchanging information in a distributed environment. SOAP provides a common message format for exchanging data between clients and services. The SOAP messages can be exchanged with the help of HTTP GET, POST, PUT requests. The HTTP GET is used to request the WSDL messages. The HTTP POST used for the service request/ response.

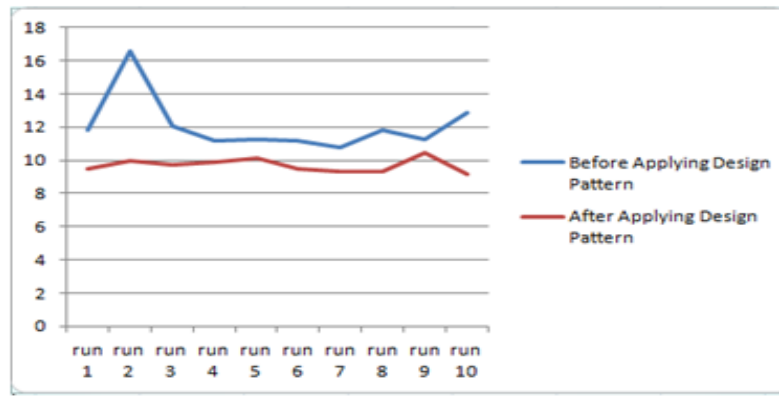

Figure 5: Profiling data for the Checklist pattern with AOP and Without AOP

## 6. PROFILING RESULTS

In order to understand the proposal in our paper let's look at an example application with SOA see Figure 2. Here initially the client system will request for the WSDL from all the service providing servers, when the client request for a complex task that need to be fulfilled by more than two service combinations. Then here as the client request for the Router, Printer and PC, but the three items are of different brands so the service providing peer server can only give the services related to different Brands of PCs, it have to contact another neighboring ServiceProvidingPeer2 server that provides the service (in our example Printer Brand) that is not available at the current requested server. At the same time to get the service of purchasing the router item requested by the same client the same peer1 server have to take the help of ServiceProvidingPeer3. Like this by the help of the other two peer systems (peer2 and peer3) the peer1 have successfully gathered the information related to the stock details and the delivery dates of the items ordered. So the ServiceProvidingPeer1 will gather all this information from the other service providing peer and give response to the client as if it is providing the three services at a time. So the composition of all the services is transparent to the PeerClient system. With the help of Web Services we can easily construct SOA.

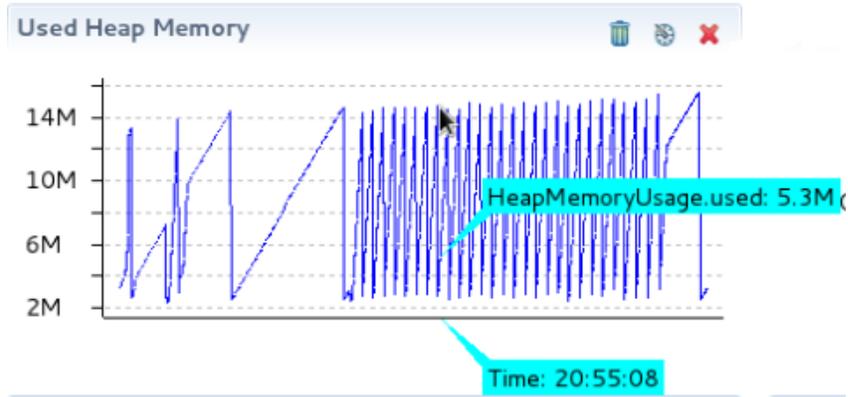

Figure 6: The Heap Memory usage after applying Aspect-Oriented Programming Techniques for implementing checklist with SOA.

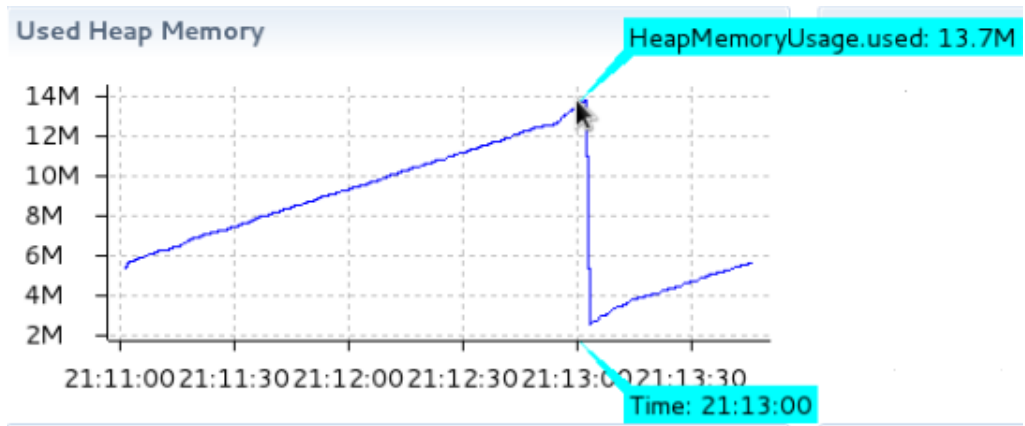

Figure 7: The Heap Memory usage after applying Object Oriented implementation of checklist design pattern with socket programming.

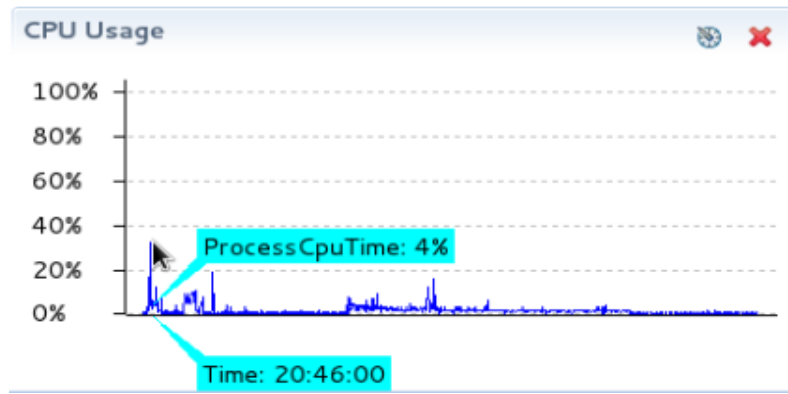

Figure 8: The CPU usage time after applying Aspect-Oriented Programming Techniques for implementing checklist with SOA.

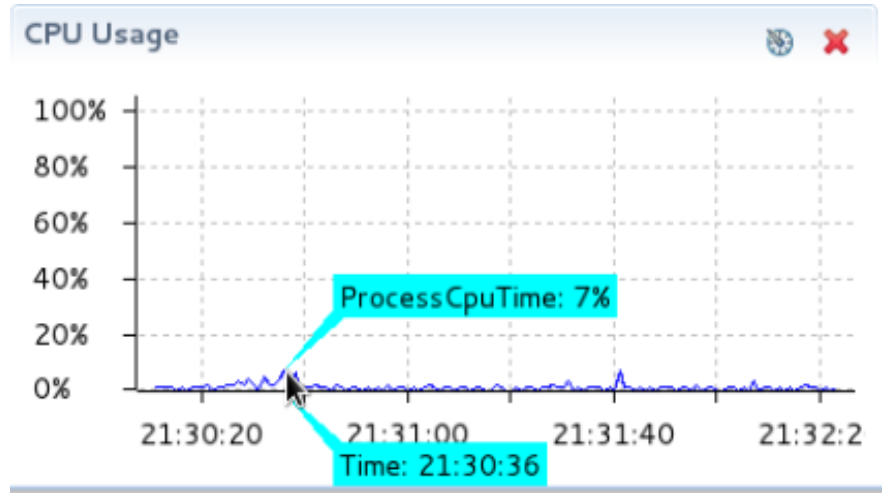

Figure 9: The CPU usage time after applying Object Oriented implementation of checklist design pattern with socket programming

The Composition of services at the service providers can be realized with the help of this proposed structure of composing the web services with SOA see Figure 2.

The view of our proposed design pattern for the unstructured peer-to-peer computing System can be seen in the form of a class diagram see Figure 3.

The flow of control in the unstructured peer-to-peer computing System can be shown with a sequence diagram in Figure 4.

We are presenting the profiling results taken for ten runs without applying this pattern and after applying this pattern using the profiling facility available in the Netbeans IDE. The graph is plotted taking the time of execution in milliseconds on Y-axis and the run count on the X-axis. The graph has shown good results while executing the code with patterns and is shown in Figure 5.This can confirm the efficiency of the proposed pattern.

### 6.1. Simulation Results after applying our Checklist oriented pattern

In our implementation we can evaluate the effectiveness of our implemented case-study with SOA. In order to make our proposal clear we have successfully developed some critical parts of our system i.e., implementation of checklist pattern integrated with SOA with the help of Aspect-Oriented Programming (AOP) and new service injection as a Feature with the help of Feature-Oriented Programming (FOP), and at the same time we have implemented the same pattern with Object-Oriented programming with pure socket programming.

The simulation results for the code developed to prove the benefits of Aspect Oriented Programming features are collected with respect to:

- Used Heap memory
- Process CPU Time

### 6.2. Discussion

From the Figure 6 and Figure 7 we can evaluate that the amount of Heap Memory used by applying Object Oriented Programming to Checklist pattern integrated with socket programming is 50,393 Kbytes and where as for the amount of Heap Memory used with AOP based Checklist pattern along with SOA is 21,302 Kbytes. It's clear that the application developed using Aspect-Oriented Programming along with SOA takes less heap memory when compared to implementation with respect to Object Oriented checklist design pattern.

From the Figures 8 and Figure 9 we can evaluate that the amount of CPU Time used with AOP based Checklist pattern along with SOA is 6.450sec and where as for the amount of CPU Time used by applying Object Oriented Programming to Checklist pattern integrated with socket programming is 12.392 sec. It's clear that the application developed using Aspect-Oriented Programming takes less CPU Time when compared to implementation with respect to OOP based checklist design pattern.

## 7. CONCLUSION AND FUTURE WORK

In this paper we have proposed a pattern to facilitate the ease of developing unstructured peer-to-peer computing systems. So with the help of our proposed design pattern named Service Injection Design pattern for unstructured peer-to-peer systems, provide services to clients with the help of Aspect-Oriented design patterns. So with this pattern we can handle the service-request of the clients and inject the new services into peer's code as feature modules. Several future directions of work are possible. We are examining how these design patterns can be inserted into a middleware technology , so that we can provide the Autonomic properties inside the middleware technologies like JXTA with the help of Features-Oriented and Aspect-Oriented programming methods.

International Journal on Web Service Computing (IJWSC), Vol.3, No.3, September 2012

## A. INTERFACES DEFINITION FOR THE PATTERN ENTITIES

Some of the Interfaces for the classes are provided as below:

**Service Providing Peer**
```
Public class Service Providing Peer{
Public Socket Server(){
}
Public static Pipe Advertisement create Socket Advertisement(){
}
Public void run(){
}
}
```

**Connection Handler:**
```
private class ConnectionHandler implements Runnable
{
Socket socket = null;
Public ConnectionHandler(){
}
private void sendAndReceiveData(Socket socket) {
}
Run(){
 }}
```

**Aspect Worker Object**
```
Public Aspect Worker Object{
Public pointcut workeroper():execution(. . )||execution(. .);
Void around(): workeroper(){
Runnable worker=new Runnable(){
 }
 }
 }
```

**ServiceRepository**
```
Public class ServiceRepository{
Public DataBaseofServices(){
}
}
```